\documentclass[aps,showpacs,preprintnumbers,amsmath,amssymb]{revtex4}
\usepackage{dcolumn}
\usepackage[dvips]{epsfig}
\def \be {\begin{equation}}
\def \ee {\end{equation}}
\def \bea {\begin{eqnarray}}
\def \eea {\end{eqnarray}}

\begin{document}
\baselineskip=0.8 cm
\title{Holographic insulator/superconductor transition with exponential nonlinear electrodynamics probed by entanglement entropy}
\author{Weiping Yao$^{a}$, Chaohui Yang$^{a}$, Jiliang Jing$^{b}$\footnote{Corresponding author, Email: jljing@hunnu.edu.cn}}
\affiliation{$^{a}$Department of electrical engineering, Liupanshui Normal University, Liupanshui, Guizhou 553004, P. R. China.}

\affiliation{$^{b}$Department of Physics, Key Laboratory of Low Dimensional \\ Quantum Structures and Quantum Control of Ministry of Education, and Synergetic Innovation Center for Quantum Effects and Applications, Hunan Normal University, Changsha, Hunan 410081, P. R. China.}

\begin{abstract}
\baselineskip=0.6 cm
\begin{center}
{\bf Abstract}
\end{center}
From the viewpoint of holography, we study the behaviors of the entanglement entropy in insulator/superconductor transition with exponential nonlinear electrodynamics (ENE). We find that
the entanglement entropy is a good probe to the properties of the holographic phase transition.
Both in the half space and the belt space, the non-monotonic behavior of the entanglement entropy in superconducting phase versus the chemical potential is general in this model. Furthermore,
the behavior of the entanglement entropy for the strip geometry shows that the confinement/deconfinement phase transition appears in both insulator and superconductor phases. And
the critical width of the the confinement/deconfinement phase transition depends on the chemical potential and the exponential coupling term. More interestingly, the behaviors of the entanglement entropy
in their corresponding insulator phases are independent of the exponential coupling factor but depends on the the width of the subsystem $\mathcal{A}$.
\end{abstract}

\maketitle

\newpage

\section{Introduction}
As a strong-week duality, the anti-de Sitter/conformal field theories (AdS/CFT) correspondence
\cite{J.Maldacena1998, EWritten1998, S.S.Gubser1998} establishes a dual relationship
between the $(d-1)$ dimensional strongly interacting theories on the boundary and the
$d$ dimensional weekly coupled gravity theories in the bulk. Based on this novel idea,
the AdS/CFT correspondence have received considerable interest in modeling strongly
coupled physics, in particular the construction of the holographic superconductor,
might shed some light on the problem of understanding the mechanism of the high temperature superconductors
in condensed matter physics \cite{
Hartnoll-PRl, Hartnoll2008, Horowitz2008, S.A.Hatnall2009, Horowitz2010, G.T.Horowitz2010}. Such holographic superconductor models are interesting
since they exhibit many characteristic properties shared by real superconductor.
In recent years, the studies on the holographic superconductors in AdS spacetime
have received a lot of attentions\cite{
Jing2010, peng2011, Gangopadhyay2012, xixu zhao2012, Gangopadhyay2012-1, Jiliang2012, Yao2013,
Dey2015, Lai2015, Hamid2016, Zeinab2017, Sheykhi2017}.

In addition, the entanglement entropy is expected to be a useful tool
to keep track of the degrees of freedom of strongly coupled systems while other traditional methods might
not be available. In the spirit of AdS/CFT correspondence, a holographic method for calculating
the entanglement entropy has been proposed by Ryu and Takayanagi\cite{Ryu:2006bv, Ryu:2006ef}.
Presently, consider a subsystem $\mathcal{A}$  of the total boundary system, the entanglement entropy for a region $\mathcal{A}$  of the boundary system is obtained from gravity side as the area of the minimal surface $\gamma_\mathcal{A}$ in the bulk which ends at $\partial\mathcal{A}$. Then the entanglement entropy
of $\mathcal{A}$ with its complement is given by
\begin{equation}\label{law}
S_\mathcal{A}=\frac{\rm Area(\gamma_\mathcal{A})}{4G_N},
\end{equation}
where $G_N$ is the Newton's constant in the bulk.
With this elegant and executable approach, the holographic entanglement entropy
has recently been applied to disclose properties of phase transitions in various holographic
superconductor models \cite{Fursaev2006, Hirata2007, Nishioka2007, Klebanov2007, Myers:2012ed, Pakman2008,
Albash2012-1, Cai2012, Cai2013, deBoer:2011wk, Hung:2011xb, Ogawa2011}. It turns out that the entanglement entropy is a good probe to the critical phase transition points and the order of holographic phase transition\cite{ Xi Dong2013,
Kuang2014, Peng2014, Zeng2016, Mazhari2016, Peng2017}.
However, most studies on the holographic entanglement entropy are focused on the cases where the gauge
field is in the form of the Maxwell field. When thinking about the higher derivative correction to the gauge field, the Refs.\cite{Yao2014,Yao2014-1} studied the holographic entanglement entropy in superconductor transition with Born-Infeld electrodynamics. Then, the behaviors of holographic entanglement entropy in the time-dependent background with nonlinear electrodynamics has been present in \cite{Liu2016}.

In 1930's Born and Infeld \cite{Born} introduced the theory of nonlinear electrodynamics to avoid the infinite self energies for charged point particles arising in Maxwell theory.   The ENE theory, as a extended Born-Infeld-like nonlinear electrodynamics,  was introduced by Hendi \cite{Hendi2012, Hendi2013}. It's Lagrangian density is $L=\frac{1}{4b^2}\bigg[e^{-b^2 F^2}-1\bigg]$ with
$F^2=F^{\mu\nu}F_{\mu\nu}$. When the ENE factor $b\rightarrow 0$, the Lagrangian will
reduce to the Maxwell case. Compared to the Born-Infeld nonlinear electrodynamics (BINE),
the ENE displays different effect on the electric potential and
temperature for the same parameters and its singularity is much weaker than the Einstein-Maxwell theory \cite{Hendi2014, Hendi2015}.
Recently, this theory has applications in several branches of physics being particularly interesting in systems where the ENE is minimally coupled with gravitation as in the cases of
charged black holes \cite{A.Sheykhi, A.Sheykhi2014, Kord Zangeneh2016, Kruglov2016, Hajkhalili2018} and cosmology \cite{Lorenci2002, Novello2004, Novello2009}.

Consequently,
it is of great interest to investigate the holographic entanglement entropy in AdS spacetime by considering the exponential form of nonlinear electrodynamics. In our previous work \cite{Yao2016}, we have investigated the effects of the ENE sector on the holographic entanglement entropy in metal/superconductor phase transition. As a further step along this line, in this paper, we will further study the properties of phase transitions by calculating the behaviors of the scalar operator and the entanglement entropy in holographic insulator/superconductor model with ENE.

The paper is organized as follows. In the next section, we will derive the equations of motions
and give the boundary conditions of the holographic model in AdS soliton spacetime. Then in section III,
we will study the properties of holographic phase transition by examining
the scalar operator. In Section IV, we will calculate the holographic entanglement entropy in insulator/superconductor transition with ENE. Finally, Section V is devoted to conclusions.

\section{Equations of motion and boundary conditions}
The action for a ENE field coupling
with a charged scalar field with a
negative cosmological constant in five-dimensional spacetime reads
\begin{eqnarray}\label{action}
S&=&\int d^5
x\sqrt{-g}\left[R+\frac{12}{L^2}-|\nabla\Psi-iqA\Psi
|^2-m^2|\Psi|^2+\frac{1}{4b^2}\left(e^{-b^2F}-1\right)\right],
\end{eqnarray}
where $g$ is the determinant of the metric,
$L$ is the radius of AdS spacetime, $q$ and $m$ are respectively the
charge and the mass of the scalar field, $F=F_{\mu\nu}F^{\mu\nu}$ here $F_{\mu\nu}$ is the electromagnetic field tensor. The Einstein equation derived from the above action becomes
\begin{equation}\label{eq-Einstein}
  R_{\mu\nu}-\frac{1}{2}g_{\mu\nu}R-\frac{6}{L^2}g_{\mu\nu}=\frac{1}{2}T_{\mu\nu},
\end{equation}
where the energy-momentum tensor $T_{\mu\nu}$ is
\begin{eqnarray}\label{eq-energy-momentum tensor}
T_{\mu\nu}&=&e^{-b^2F}F_\mu\lambda F^\lambda_\mu+2\nabla_\mu\psi\nabla_\nu\psi+2q^2A_\mu A_\mu \psi^2
\nonumber \\
&+&g_{\mu\nu}\left[\frac{1}{4b^2}\left(e^{-b^2F}-1\right)
-\nabla_\mu\psi\nabla^\nu\psi-q^2A_\mu A^\nu \psi^2-m^2\psi^2\right].
\end{eqnarray}
The equations of the motion of the matter fields can be obtained in the form
\begin{eqnarray}\label{eq-matter field}
&& \frac{1}{\sqrt{-g}}\partial_{\mu}(\sqrt{-g}g^{\mu\nu}\partial_\nu\psi)-q^2A_{\mu}A^{\mu}\psi-m^2\psi=0,\\
&&\frac{1}{\sqrt{-g}}\partial_{\mu}(\sqrt{-g}e^{-b^2F}F^{\mu\nu})-2q^2A^\nu\psi^2=0.
\end{eqnarray}
For simplicity, our ansatz for the metric and matter fields are given by
\begin{eqnarray}
&&ds^2 =\frac{dr^2}{r^2B(r)}+r^2\left(-e^{C(r)}dt^2+dx^2+dy^2+e^{A(r)}B(r)d\chi^2\right),\\
&&A_t=\phi(r),\ \ \ \ \psi=\psi(r).
\end{eqnarray}
Without lose of generality, we set $L=1$ in this paper. In order to
get a smooth geometry at the tip $r_0$ satisfying $B(r_0)=0$, $\chi$ should be
made with an identification
\begin{equation}
\chi=\chi+ \Gamma,\ \ \ \  with \ \  \Gamma=\frac{4\pi e^{-A(r_0)/2}}{r_0^2B'(r_0)}\;.
\end{equation}
The independent equations of motion under the above ansatz can be obtained as follows
\begin{eqnarray}
&&
\psi''+\left(\frac{5}{r}+\frac{A'}{2}+\frac{B'}{B}+\frac{C'}{2}\right)\psi'+\frac{1}{r^2B}
\left(\frac{e^{-C}q^2\phi^2}{r^2}-m^2\right)\psi=0\;,
 \\
&&(1+4b^2e^{-C}B\phi'^2) \phi''+\left(\frac{3}{r}+\frac{A'}{2}+\frac{B'}{B}
-\frac{C'}{2}\right)\phi'+2b^2e^{-C}B\left(\frac{B'}{B}-C'\right)\phi'^3\nonumber \\ &&\ \ \ \
-\frac{2e^{-2b^2e^{-C}B\phi'^2}q^2\psi^2\phi}{r^2B}=0\;,
\\
&& A'=\frac{2r^2C''+r^2C'^2+4rC'+4r^2\psi'^2-2e^{-C+2b^2e^{-C}B\phi'^2}\phi'^2}{r(6+rC')}\;, \label{Aeq}
 \\
 &&
C''+\frac{1}{2}C'^2+\left(\frac{5}{r}+\frac{A'}{2}+\frac{B'}{B}\right)C'
-\frac{e^{-c}}{r^2}\left(e^{2b^2e^{-C}B\phi'^2}\phi'^2+\frac{2q^2\phi^2\psi^2}{r^2B}\right) =0,
\\
&&
B'\left(\frac{3}{r}-\frac{C'}{2}\right)+B\left[\psi'^2-\frac{1}{2}
A'C'+\frac{e^{-C+2b^2e^{-C}B\phi'^2}\phi'^2}{r^2}+\frac{12}{r^2}
\right] \nonumber \\
&&\ \ \ \ +\frac{1}{r^2}\left(\frac{e^{-C}q^2\phi^2\psi^2}{r^2}+\frac{1-e^{2b^2e^{-C}B\phi'^2}}{4b^2}
+m^2\psi^2-12\right)
=0,
\end{eqnarray}
where the prime denotes the derivative with respect to $r$.
 For the sake of integrating
the field equations from the tip of the soliton out to the infinity
for this system, we need
to specify the asymptotic behavior both at the tip and the infinity.
At the tip $(r=r_0)$, the above
equations can be Taylor expand in the form \cite{peng2011}
\begin{eqnarray}
&&\psi(r)=\psi_0+\psi_1(r-r_0)+...,\nonumber \\
&&\phi(r)=\phi_0+\phi_1(r-r_0)+...,\nonumber \\
&&A(r)=A_0+A_1(r-r_0)+...,\nonumber \\
&&B(r)=B_0(r-r_0)+B_1(r-r_0)^2+...,\nonumber \\
&&C(r)=C_0+C_1(r-r_0)+....
\end{eqnarray}
The boundary conditions near the AdS boundary where $r\rightarrow \infty$ are
\begin{eqnarray}
&&\psi\sim\frac{\psi_{-}}{r^{\Delta_{-}}}+\frac{\psi_{+}}{r^{\Delta_{+}}},\ \ \ \
\phi\sim\mu-\frac{\rho}{r^{2}}\label{b:infinityphi},\nonumber \\
&&A\sim\frac{A_4}{r^4}+...,\ \ \ \
B\sim 1+\frac{B_4}{r^4}+...,\ \ \ \
C\sim\frac{C_4}{r^4}+....
\end{eqnarray}
Where the conformal dimensions of the
operators are $\Delta_{\pm}=2\pm\sqrt{4+m^2}$, $\mu$ and $\rho$ can be interpreted as the
the corresponding chemical potential and
charge density in the dual field theory respectively.
According to the AdS/CFT
correspondence, both $\psi_-$ and $\psi_+$ can be normalizable and
they correspond to the vacuum expectation values
$\psi_{-}=<\mathcal{O}_{-}>$,  $\psi_{+}=<\mathcal{O}_{+}>$ of
an operator $\mathcal{O}$ dual to the scalar field\cite{Hartnoll-PRl, Hartnoll2008}.
Further, the above equations of
motion have useful scaling symmetries\cite{G.T.Horowitz2010}
\begin{eqnarray}
r\rightarrow \alpha r,\qquad (\chi,x,y,t)\rightarrow(\chi,x,y,t)/\alpha,\qquad\phi\rightarrow \alpha\phi\label{scaling1},
\end{eqnarray}
Using the scaling symmetries~(\ref{scaling1}), we can take $r_0=1$.
 And the useful quantities can be scaled as
\begin{equation}\label{scaling3}
\Gamma\rightarrow\frac{1}{\alpha}\Gamma,\ \ \mu\rightarrow\alpha\mu,
\ \ \rho\rightarrow\alpha^3\rho,\ \ \langle\hat{O}_+\rangle\rightarrow\alpha^{\frac{5}{2}}\langle\hat{O}_+\rangle.
\end{equation}
Therefore, we will use the following dimensionless quantities in next section
\begin{equation}\label{scaling33}
\mu\Gamma,\ \ \ \ \rho\Gamma^3, \ \ \ \ \langle\hat{O}_+\rangle^{\frac{2}{5}}\Gamma,
\end{equation}

\section{Insulator/Superconductor Phase Transition}
In this section, we want to study of the phase transition in the
five-dimensional AdS soliton background with ENE field.
In order to obtain the the solutions in the complicated model and ensure the validity of the results, we here concentrate on the case in the weak effects of ENE field and study its influences on the properties of the holographic phase transition.
From above discussion, for given $m^2, q, \psi{(r_0)}$, we can solve the equations of motion
by choosing $\phi(r_0)$ as a shooting parameter.  Considering the BF bound\cite{Breitenlohner 1982},
we choose $m^2=-\frac{15}{4}, q=2$ in this paper. Then, $\psi_-$ can either be identified
as an expectation value or a source of the operator $\mathcal{O}$ of the dual superconductor.
In the following calculation, we will consider $\psi_-$ as the source of the operator and use
the $\psi_{+}=<\mathcal{O}_{+}>$ to describe the phase transition in the dual CFT.\ \
\begin{figure}[ht]
\centering
\includegraphics[scale=0.42]{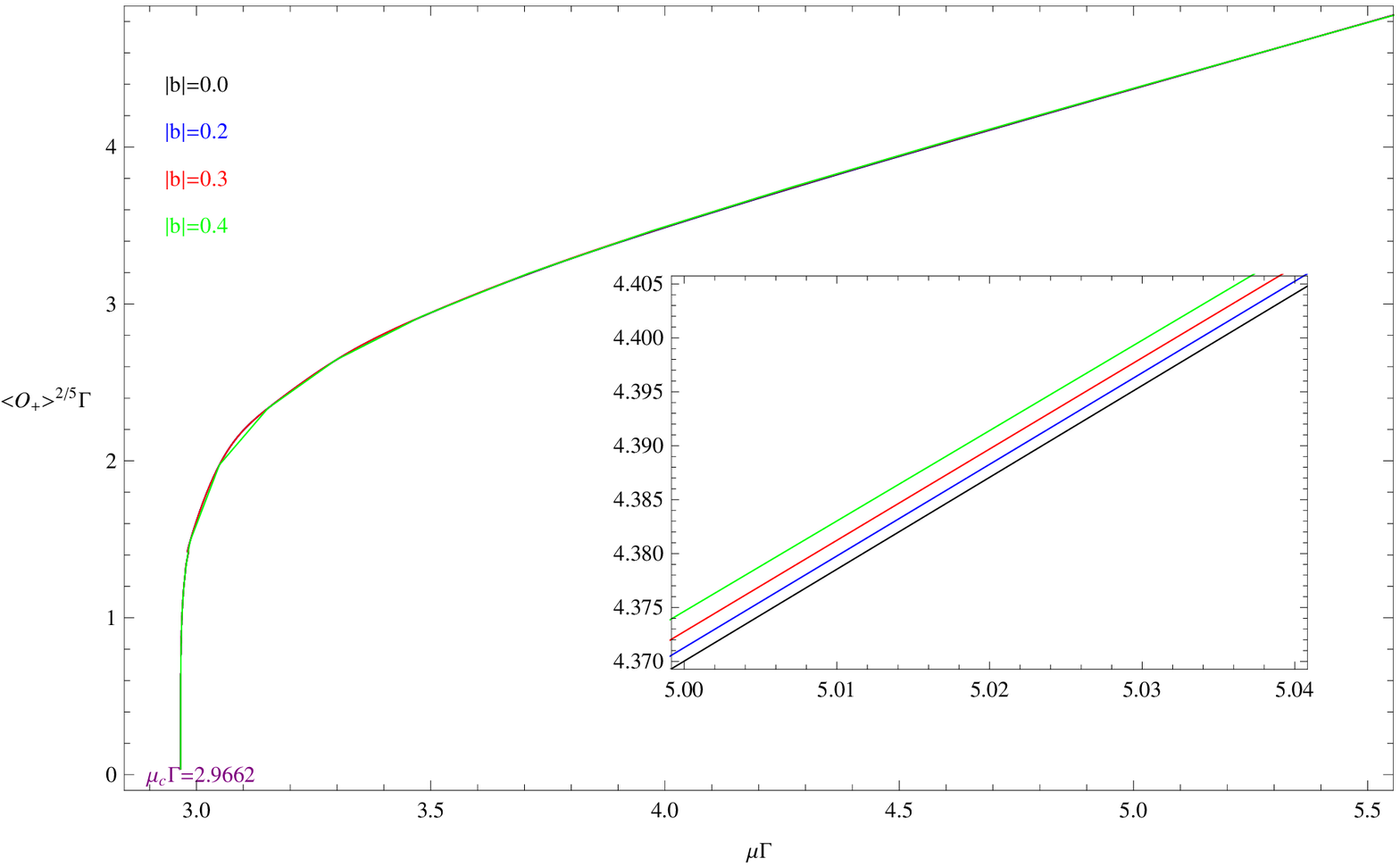}\hspace{0.2cm}%
\includegraphics[scale=0.66]{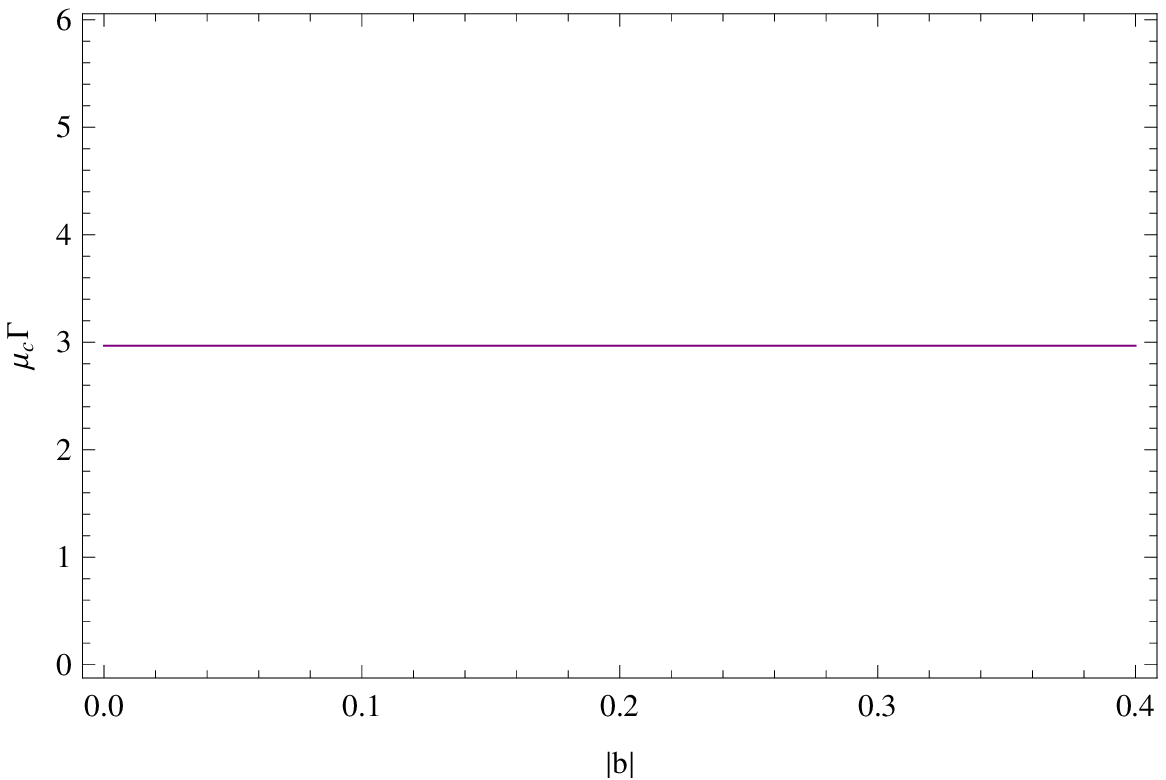}
\caption{\label{condesation} The condensate of operator
 $\langle\hat{O}_+\rangle$ (left plot) as a function of chemical
 potential $\mu$ for different parameter $b$ and the critical chemical potential $\mu_c$
 (right plot) versus the ENE factor $b$. The different color correspond to the different
 value of parameter $b$.}
\end{figure}
\begin{figure}[ht]
\centering
\includegraphics[scale=0.58]{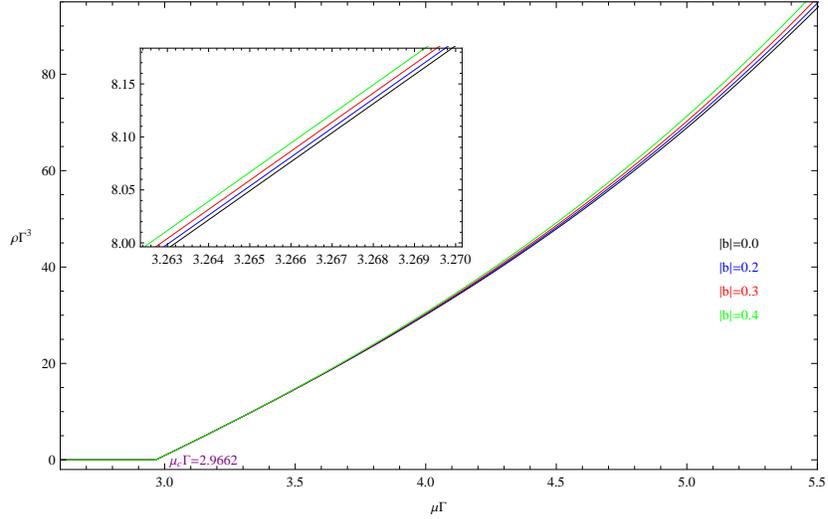}
\caption{\label{charge}The charge density $\rho$ as a function of chemical
 potential $\mu$ for different parameter $b$, respectively. The enlarged view in the red elliptic region is
 plotted in the right panel. The different
 color correspond to the different value of parameter $b$. }
\end{figure}

Here, we plot pictures to display the explicit dependence of the chemical potential for operator $\langle\hat{O}_+\rangle$ and charge density $\rho$ on the ENE factor $b$.
It can be seen from the Fig. \ref{condesation} that there is a phase transition at the
critical chemical potential $\mu_c$ and its value is independent of the ENE factor b which is shown in the
right-hand panel. That is to say, the ENE has no effect on the critical potential of the holographic phase transition for this physical model. When $\mu<\mu_c$, the system is described by the AdS soliton solution itself
which indicates a insulator phase turns on. When the chemical potential is bigger than the critical value
$\mu_c$ the condensation of the operator emerges, which means the AdS soliton reaches a superconductor phase.
 It is interesting to find that the effect of the ENE factor $b$ on the operator in the condensate phase is not trivial. With the increase of the strength of the ENE, the value of the scalar operator becomes bigger.
In Fig. \ref{charge}, we note that the charge density $\rho$ in the superconductor phase drops when the ENE parameter becomes lower and the insulator/superconductor phase transition here is typically the second order in this case.

\section{Holographic Entanglement Entropy}
In this section, we will study the behavior of holographic entanglement entropy
in this holographic model and discuss the effect of the
ENE factor $b$ on the entanglement entropy. Since the choice of the subsystem
$\mathcal{A}$ is arbitrary, we can define infinite entanglement entropy correspondingly.
For concreteness, we investigate the holographic entanglement entropy of dual field with a half space
and a belt geometry in the AdS boundary, respectively.
\subsection{Holographic Entanglement Entropy for half geometry}
\begin{figure}[ht]
\centering
\includegraphics[scale=0.80]{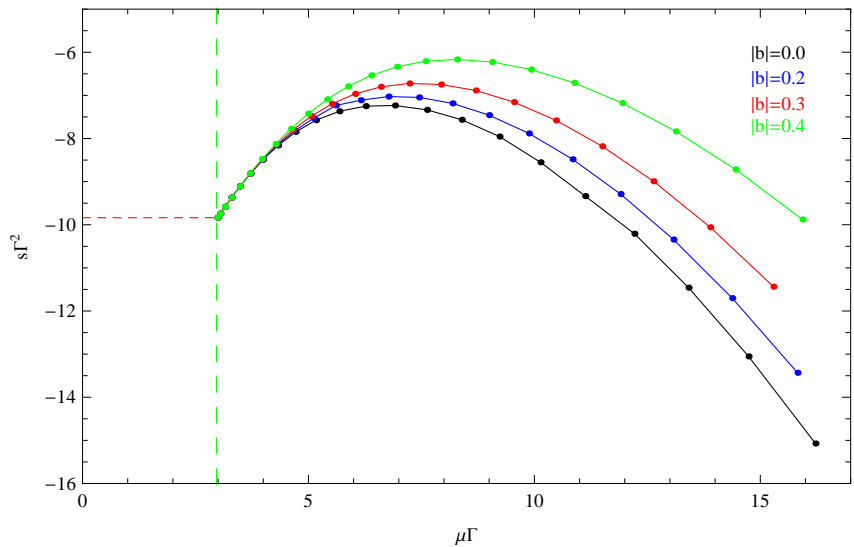}
\caption{\label{halfHEE} The entanglement entropy $s$ as a function of the chemical potential
for different ENE factor $b$ in the half geometry. The red horizontal dashed line denotes the entropy in the insulator phase, the green vertical dashed line denotes the critical phase transition point where $\mu_c\Gamma=2.9662$ ,and the solid curves denote the entropy in the superconductor phase for various
 ENE factors: black curve for $\vert b\vert=0$, blue curve for $\vert b\vert=0.2$.
 red curve for $\vert b\vert=0.3$ and  green curve for $\vert b\vert=0.4$.}
\end{figure}

We first consider the subsystem $\mathcal{A}$ with a half space defined by $x>0$,\
 $-\frac{R}{2}<y<\frac{R}{2}(R\rightarrow\infty),\ 0\leq\chi\leq\Gamma$.
According to the proposal (\ref{law}), the  entanglement entropy can be expressed as
\begin{equation}\label{half1}
S_\mathcal{A}^{half}=\frac{R\Gamma}{4G_N}\int_{r_0}^{\frac{1}{\epsilon}}re^{\frac{A(r)}{2}}dr=\frac{R\pi}{8G_N}(\frac{1}{\epsilon^2}+s),
\end{equation}
where $r=\frac{1}{\epsilon}$ is the UV cutoff. Note that the first term is divergent
as $\epsilon\rightarrow0$ and will not change after the operator condensation\cite{Cai2012}.
The second term $s$ is independent of the UV cutoff and $s=-1$ corresponds to the pure AdS
soliton. As the aim of requiring the the lower bound of the integral is still equal to unit, we define a useful dimensionless coordinate in the form
\begin{equation}
z=\frac{r_0}{r}.
\end{equation}
Then, the entanglement entropy for a half space can be rewritten as
\begin{equation}\label{half2}
S_\mathcal{A}^{half}=-\int_{1}^{\epsilon r_0}\frac{r_0^2e^{\frac{A(z)}{2}}}{z^3}dz=\frac{1}{2}(\frac{1}{\epsilon^2}+s),
\end{equation}
while the second term $s$ is a finite term, so it is physically important. According to the
scaling symmetry (\ref{scaling1}), we here choose the following scale invariants to
explore physics in the entanglement entropy $s$.
\begin{equation}
s\Gamma^2, \ \ \ \mu\Gamma^2.
\end{equation}
In Fig. \ref{halfHEE}, we plot the behavior of the entanglement entropy $s$  with respect of
chemical potential $\mu$ and the ENE factor $b$ in the half geometry. It can be seen from
the figure that the entanglement entropy is continuous but its slop has a discontinuous change at the critical
phase transition point $\mu_c$. Which indicates some kind of new degree of freedom like the Cooper pair
would emerge after the condensation. Furthermore, the discontinuous change of the entanglement entropy at $\mu_c$ signals that the phase transition here is the second order transition. With the increase of the ENE
factor, the value of $\mu_c$ dose not change. Which means the ENE parameter has no effect on the critical point of the phase transition. Before the phase transition, the entanglement entropy is a constant as we change the parameters $b$ and $\mu$ which can be interpreted as the insulator phase. After the phase transition,
for a given $b$, the entanglement entropy in the superconductor phase first increases and then decreases monotonously for larger $\mu$. And the value of the entanglement entropy
becomes lager as we choose a lager ENE parameter for a given $\mu$.
When the factor $b\rightarrow0$, the ENE field will reduce to the Maxwell field and our results are
 consistent with the one discussed in Ref.\cite{Cai2012}.

\subsection{Holographic Entanglement Entropy for strip  geometry}
\begin{figure}[ht]
\centering
\includegraphics[scale=0.66]{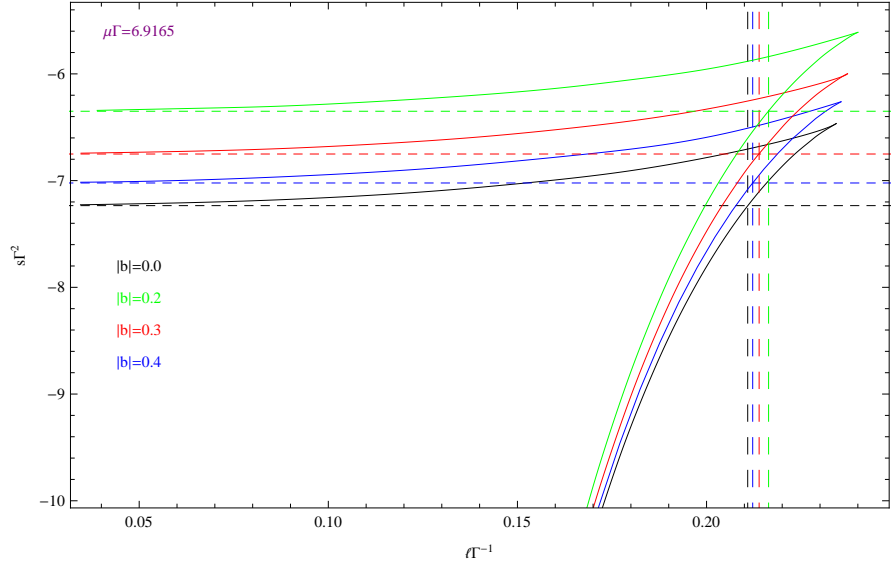}
\caption{\label{Entropy} The entanglement entropy as a function of the
strip width $\ell$ with different parameter $b$ for $\mu\Gamma=6.9165$.
The black curve is for $\vert b\vert=0$ and the blue curve is for $\vert b\vert=0.2$.
The red curve is for $\vert b\vert=0.3$ and the green curve is for $\vert b\vert=0.4$.
The corresponding critical widths are $\ell_c\Gamma^{-1}=0.2109$, $\ell_c\Gamma^{-1}=0.2122$, $\ell_c\Gamma^{-1}=0.2139$, $\ell_c\Gamma^{-1}=0.2164$, respectively.}
\end{figure}

In the Following calculation, we are interested in a more nontrivial geometry which is a strip shape for region $\mathcal{A}$. We assume that the strip shape with a finite width $\ell$ along the x direction, along the $\eta$ direction with a period $\Gamma$, but infinitely extending in $y$ direction.
The holographic dual surface $\gamma_\mathcal{A}$ defined as a codimension three surface is
$t=0,\ \ x=x(r),\ \ -\frac{R}{2}<y<\frac{R}{2}\ (R\rightarrow\infty),\ \ 0\leq\chi\leq\Gamma$.
 Considering the surface is smooth, we suppose that
the holographic surface $\gamma_\mathcal{A}$ starts from
$x=\frac{\ell}{2}$ at $r=\frac{1}{\epsilon}$, extends into the bulk until it reaches $r=r_*$,
then returns back to the AdS boundary $r=\frac{1}{\epsilon}$ at $x=-\frac{\ell}{2}$.
The entanglement entropy of the belt geometry with connected surface in $z$ coordinate is given by
\begin{equation}\label{surface}
S^{connect}_\mathcal{A}[x]=-\frac{R\Gamma}{2G_N}\int_{z*}^{r_0\epsilon}\frac{r_0^2}{z^3}e^{\frac{A(z)}{2}}
\sqrt{1+r_0^2B(z)(dx/dz)^2}dz,
\end{equation}
where $z_*=r_0/r_*$ is the turning point. As the physics model is a static situation, the above
Eq.(\ref{surface}) dose not depend on the time slice. And we could consider $S^{connect}_\mathcal{A}[x]$
as the integral of the Lagrangian with x direction thought of as time. Because the translations of $x$ direction
is symmetry, the corresponding Hamiltonian is conserved. Therefore, the equation of motion for the minimal surface from Eq.(\ref{surface}) can be deduced as following
\begin{equation}\label{minimal}
\frac{r_0^4 B(z)(dx/dz)e^{\frac{A(r)}{2}}}{z^3\sqrt{1+r_0^2B(z)(dx/dz)^2}}
=\frac{r_0^3 \sqrt{B(z_s)}e^{\frac{A(z_s)}{2}}}{z_s^3},
\end{equation}
where $z_s$ is a constant. And the width $\ell$ and the entanglement entropy $S$ can be
easily calculated in the form
\begin{eqnarray}
\frac{\ell}{2}&=-&\int_{z_{s*}}^{r_0\epsilon}\frac{1}{r_0
\sqrt{B(z)(\frac{z_*^6B(z)e^{A(z)}}{z^6B(z_*)e^{A(z_*)})}-1)}}dz_s,\\
S^{connect}_\mathcal{A}&=-&\int_{z_{s*}}^{r_0\epsilon}\frac{r_0^2}{z^3}e^\frac{A(z)}{2}
\sqrt{1-\frac{z^6B(z_*)e^{A(z_*)}}{z_*^6B(z)e^{A(z)}}}dz+\frac{\ell}{2}\sqrt{\frac{1}{z^6}B(z)e^{A(z)})}r_0^3
=\frac{1}{2}(\frac{1}{\epsilon^2}+s),
\end{eqnarray}
In addition to the solution to the connected configuration,
the entanglement entropy for the disconnected geometry described
two separated surfaces that are located at $x=\pm\frac{\ell}{2}$ respectively and extending to the bulk and reaching at the tip is given by
\begin{equation}\label{half}
S_\mathcal{A}^{disconnect}=-\frac{R\Gamma}{2G_N}\int_{1}^{ r_0\epsilon}\frac{r^2_0e^{\frac{A(z)}{2}}}{z^3}dz
=\frac{R\pi}{4G_N}(\frac{1}{\epsilon^2}+s),
\end{equation}
Here, we show in Fig. \ref{Entropy} the results for the entanglement entropy $s$
versus the width $\ell$ of the subsystem $\mathcal{A}$ and the ENE factor $b$ with the dimensionless
quantities $s\Gamma^2, \mu\Gamma, \ell\Gamma^{-1}$ and $b$. We find that the the discontinuous solutions represented the horizontal dotted lines in the figure is independent of the the width $\ell$ but its value of the entanglement entropy with a smaller $b$ is smaller. The connected configuration denoted by
the solid lines has two solutions. Specifically, the so-called confinement/deconfinement phase transition \cite{Nishioka2007, Klebanov2007, Myers:2012ed} emerges as we change the width $\ell$ and the critical value $\ell_{c}$ indicated by the vertical dotted lines becomes bigger with the increase of the parameter $b$. For a fixed $b$, in the deconfinement phase where $\ell<\ell_c$, considering the physical entropy determined by the choice of the lowest one, the
entanglement entropy comes from the connected surface and the lowest branch in the figure is
finally favored. However, the physical entanglement entropy in confinement phase where $\ell>\ell_c$ is dominated by the discontinuous surface and has nothing to do with the factor $\ell$. Thus, there exists
four phases in the dual boundary field theory, including the insulator phase, superconductor phase and their
corresponding confinement/deconfinement phases. When the parameter $\ell$ is fixed, we observe that the entanglement entropy increases as we choose
a bigger ENE factor both in the confinement and deconfinement superconducting phases.

\begin{figure}[ht]
\centering
\includegraphics[scale=0.52]{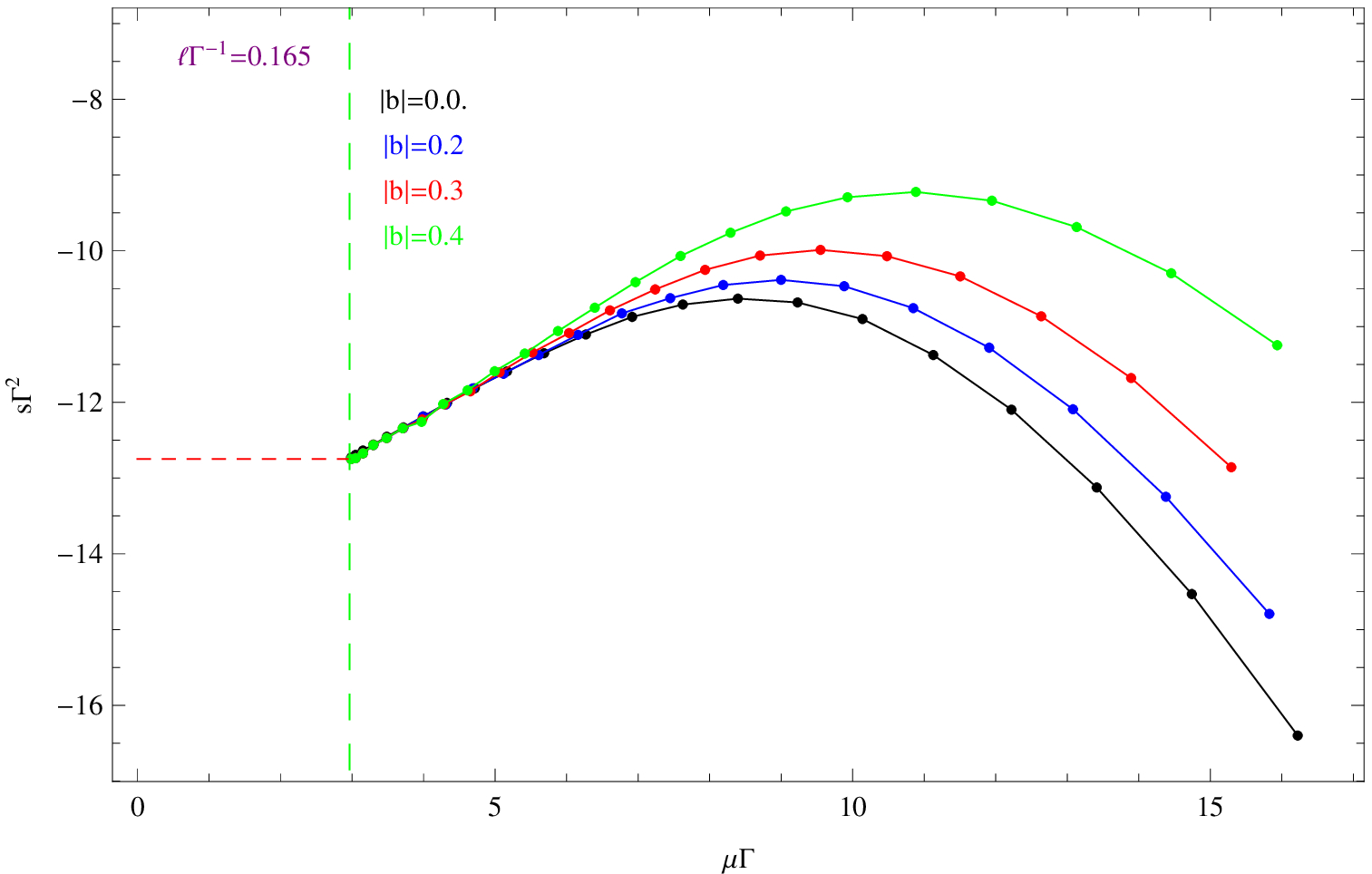}
\includegraphics[scale=0.56]{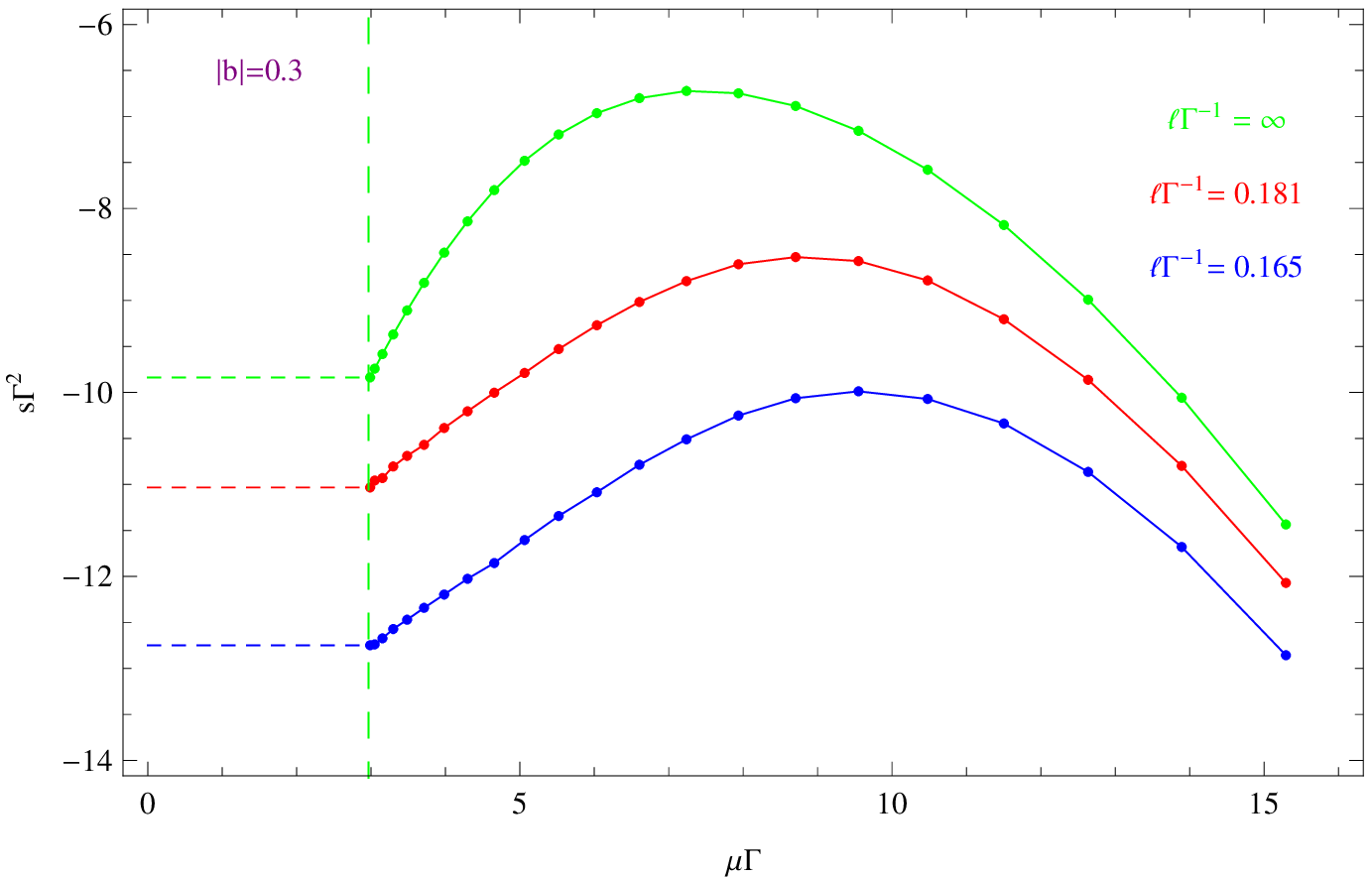}
\caption{\label{SmuL} The entanglement entropy $s$ as a function of the chemical potential $\mu$
 for the various factors,i.e., the ENE factor $b$ and the
belt width $\ell$. The horizontal dotted lines denote the entropy in the insulator phase, the green vertical dashed  line denotes the critical phase transition point where $\mu_c\Gamma=2.9662$, and the solid curves denote the entropy in the superconductor phase.
 The left-hand figure corresponds to $\ell\Gamma^{-1}=1$ and different b:
black curve for $\vert b\vert=0$ ,  blue curve for $\vert b\vert=0.2$,
red curve for $\vert b\vert=0.3$ and  green curve for $\vert b\vert=0.4$.
The right-hand figure corresponds to $\vert b\vert=0.3$ and different $\ell$:
green curve for $\ell\Gamma^{-1}=\infty$,  red curve for $\ell\Gamma^{-1}=0.181$
and blue one  for $\ell\Gamma^{-1}=0.165$.}
\end{figure}

Interestingly, the entanglement entropy with respect to the chemical
potential $\mu$ as one fixes the ENE factor $b$ or the width $\ell$ is presented in Fig. \ref{SmuL}.
At the insulator/superconductor phase transition point $\mu=\mu_c$, we also find that the
the jump of the slop of the entanglement entropy indicates that the system undergoes the second order phase transition. Both in the confinement and deconfinement superconducting phases where $\mu>\mu_c$, we can see that the behavior of the entanglement entropy as a function of the chemical potential is non-monotonic and similar to the case in the half geometry which we have discussed above. As the chemical potential is fixed,
the value the entanglement entropy becomes smaller when the factors $b$ and $\ell$ become lower.
More specifically, the the effect of the ENE factor on the entanglement entropy
is weaker than the width of the subsystem $\mathcal{A}$.
In their corresponding insulator phases where $\mu<\mu_c$, we observe that
the value of the entanglement entropy does not change
as we alter the parameter $b$ for a given $\mu$.
On the other hand, with the increase of the width the entanglement entropy increases.
That is to say, the behavior of the entanglement entropy
in the corresponding insulator phases is independent of the the ENE factor
but depends on the the width of the subsystem $\mathcal{A}$.

\subsection{phase diagram}
Finally, according to our calculation for
entanglement entropy in holographic insulator/superconductor transition with ENE field, we use a picture to display the phase diagram of
entanglement entropy with a straight geometry.
\begin{figure}[ht]
\centering
\includegraphics[scale=0.68]{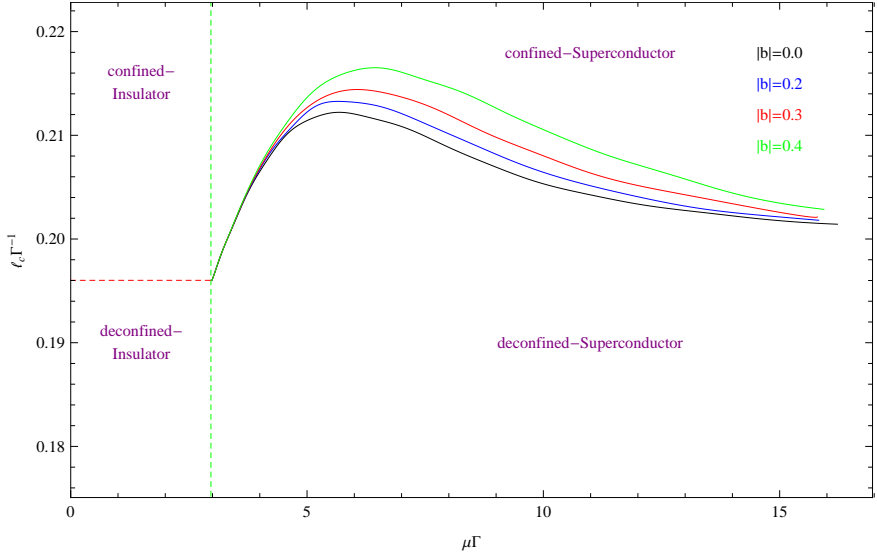}
\caption{\label{phase} The phase diagram of the entanglement entropy for a strap geometry in the
 holographic insulator/superconductor transition with ENE field.
 The black curve is for $\vert b\vert=0$ and the blue curve is for $\vert b\vert=0.2$.
The red curve is for $\vert b\vert=0.3$ and the green curve is for $\vert b\vert=0.4$. }
\end{figure}

In Fig. \ref{phase},
the insulator phase and the superconductor phase are separated by the  green vertical dashed line and the
phase boundary between the confinement phase and the deconfinement phase is separated by the red horizontal dashed line
and the solid curve. Therefore, the phases characterized by the parameters $\mu$ and $\ell$ contain the insulator phase, superconductor phase and their corresponding confinement/deconfinement phases.
It can be clearly seen from the figure that the critical width $\ell_c$ of the confinement/deconfinement phase transition in the insulator phase is independent of the ENE parameter. In the superconductor phase, however, the critical width $\ell_c$ increases with the increase of the ENE factor.
To further study, we observe that the critical width $\ell_c$ has a non-monotonic change as the chemical potential becomes bigger. Concretely, the entanglement entropy
first increases beyond the cusp at the certain chemical potential, reaches to a maximum and decreases to a
minimum, and then approaches a plateau at very large $\mu$.

\section{Summary}
We have studied the properties of phase transitions by calculating the behaviors of the scalar operator
and the entanglement entropy in holographic insulator/superconductor model with ENE.
On the basis of the behaviors of the scalar operator in this holographic model, we find
that there is a insulator/superconductor transition at the critical chemical potential point and
the effect of the ENE factor on the scalar condensation is quite different from
those observed in the holographic metal/superconductor transition with ENE field model\cite{Yao2016}.
Specifically, in the holographic insulator/superconductor system the ENE factor does not have any effect on the critical chemical potential of the transition. These conclusions can also be understood from
the behavior of the entanglement entropy. From the Fig. \ref{halfHEE}, the discontinuity of the slop of the
entanglement entropy in half space at the critical chemical potential point signals some kind of new degree of freedom like the Cooper pair would emerge after the condensation and indicates the order of associated phase transition in the system. In Fig. \ref{SmuL}, we observed the behavior of the entanglement entropy with respect chemical potential in strip geometry at the insulator/superconductor transition point is similar to the half case. That is to say, the entanglement entropy is indeed a good tool to search for the phase transition point.

In the superconducting phase, compared to the phenomenon observed in the scalar operator, the entanglement entropy versus the chemical potential displays more rich behaviors. Both in the half space and the belt space, the non-monotonic behavior of the entanglement entropy versus the chemical potential is general in this model as the ENE parameter is fixed. For a given chemical potential, the value the entanglement entropy becomes smaller when the ENE factor or the with becomes lower. In the insulator phase, however, the behavior of entanglement entropy is independent of the the ENE parameter.

Interestingly, considering the the effect of the belt width on the entanglement entropy,
we obtained that the confinement/deconfinement phase transition appears in both insulator and superconductor phases and the complete phase diagram of the entanglement entropy with a straight geometry is presented
in Fig. \ref{phase}. It is shown that the critical width of the the confinement/deconfinement phase transition depends on the chemical potential and the ENE term.

\begin{acknowledgments}
This work was supported by the National Natural Science
Foundation of China under Grant No. 11665015, 11475061;
Guizhou Provincial Science and Technology Planning Project of
China under Grant No. qiankehejichu[2016]1134;
The talent recruitment program of Liupanshui normal university of China under Grant No. LPSSYKYJJ201508.
\end{acknowledgments}

\end{document}